\documentclass[prl,aps,twocolumn,superscriptaddress,preprintnumbers,showpacs,floatfix,amsmath,amssymb]{revtex4}

\usepackage{graphicx}
\usepackage{dcolumn}
\usepackage{bm}

\usepackage{amsmath}
\usepackage{graphicx}
\usepackage{amssymb}
\usepackage{mathrsfs}
\usepackage{bbm}
\usepackage{epsfig}

\newcommand{\la}{\label}

\newcommand{\be}{\begin{eqnarray}}
\newcommand{\ee}{\end{eqnarray}}

\begin{document}

%
\title{ On correlation between protein secondary structure, \\ backbone bond angles, and side-chain orientations
}

\author{Martin Lundgren}
\email{Martin.Lundgren@physics.uu.se}
\affiliation{Department of Physics and Astronomy, Uppsala University,
P.O. Box 803, S-75108, Uppsala, Sweden}
\author{Antti J. Niemi}
\email{Antti.Niemi@physics.uu.se}
\affiliation{
Laboratoire de Mathematiques et Physique Theorique
CNRS UMR 6083, F\'ed\'eration Denis Poisson, Universit\'e de Tours,
Parc de Grandmont, F37200, Tours, France}
\affiliation{Department of Physics and Astronomy, Uppsala University,
P.O. Box 803, S-75108, Uppsala, Sweden}

\begin{abstract}
\noindent
We investigate the fine structure of the $sp3$ hybridized covalent bond geometry  that
governs the tetrahedral architecture around the central C$_\alpha$ carbon of a protein backbone, and for this
we develop new visualization techniques to analyze high resolution X-ray structures in Protein Data Bank. 
We observe that there is a correlation between the deformations of the ideal tetrahedral symmetry
and the  local  secondary structure of the protein. We propose a universal coarse grained energy 
function to describe the ensuing side-chain geometry in terms  of the C$_\beta$ carbon orientations. 
The energy function can model the side-chain geometry with a sub-atomic precision.
As an example we  construct the  C$_\alpha$-C$_\beta$ structure 
of HP35 chicken villin  headpiece. We obtain 
a configuration that deviates less than  0.4 \.A in root-mean-square distance 
from the experimental X-ray structure.
\end{abstract}




\maketitle

\section{I: Introduction}

Protein structure validation is  based on various well tested and 
broadly accepted stereochemical paradigms. Methods such 
as {\it MolProbity} \cite{mol} and {\it Procheck} \cite{pro} and many others  
help crystallographers to find and fix potential problems during 
fitting and refinement.  Stereochemical assumptions are also instrumental to structure 
prediction packages such as {\it Rosetta} and {\it I-Tasser}  \cite{ros}.  Likewise, they form the foundation 
for parameter determination in force fields such as {\it Charmm} and {\it Amber} \cite{nat}  that  aim to describe protein dynamics
at atomic scale. 

One of the  paradigms  is the transfersability assumption. It states that stereochemical 
restraints are universal and independent of the environment. Among its consequences are
that the covalent bond geometry around the backbone C$_\alpha$ should seize 
a very precise tetrahedral  $sp3$ hybridized shape. 
For example, the  backbone 
\[
\tau_{NC} \ \equiv  \ (N-C_\alpha-C) 
\]
bond angle should oscillate around a computable average value
that  depends only on the covalent bonds  between the  C$_\alpha$ and the 
N, C, H and C$_\beta$ atoms in the trans-peptide group.  In particular, at least to
to the leading order its value should 
not  depend on the character of the secondary structure 
environment. Standard molecular dynamics force fields  explicitly assume this to be the case.
These force fields are based on a harmonic approximation where the 
bond angles $\kappa$ oscillate with energy  \cite{nat} 
\begin{equation}
E_{bond} = \sum_{\rm bonds} \omega_\kappa (\kappa - \kappa_0)^2
\label{ba}
\end{equation}
Here $\omega_\kappa$ and $\kappa_0$ are parameters that are in general amino acid dependent. 
But these parameters are presumed to be independent of the geometry of the
surrounding secondary structure. Instead, they are supposed to predict the local secondary 
structure  environment.

The enormous success that has been enjoyed by the validation methods and structure 
prediction programs  in  resolving  close to 80.000 crystallographic protein structures
that are presently in Protein Data Bank (PDB) \cite{struc} is a clear manifestation that  the various
paradigms are valid to a good precision. 
However,  with the advent of third-generation synchrotron
sources of X-rays, there is now a  small but rapidly expanding number of 
protein structures  that are resolved with an ultrahigh sub-Angstr\"om resolution.
The present, third-generation X-ray synchrotron sources such as ESRF in Grenoble 
and PETRA at DESY in Hamburg can already produce photons with wavelengths as short 
as 10 pico-meters. Thus it is in principle possible to obtain three dimensional protein structures with a 
comparable resolution. The next-generation sources of high brilliance X-ray beams 
such as the European X-Ray Free Electron Laser at 
DESY, will push protein X-ray crystallography to its extreme. These future experimental 
facilities can reach both ultra-high spatial and temporal resolutions, with a fully coherent 
peak brightness that is many orders of magnitude higher than what can be obtained 
with the present third-generation synchrotron sources.  
The on-going experimental revolution in combination with the ever expanding need of 
higher precision for example in the study of protein-protein interactions, enzyme catalysis 
and search of causes for protein misfolding related diseases, are good incentives for us to 
scrutinize the level of precision in some of the paradigm assumptions on protein backbone
geometry. And, if need be, to try and develop new theoretical concepts that 
aim to describe proteins at a precision that matches the highest present and near future 
experimental standards, in revealing the finer structures of folded proteins.

In fact, {\it ab initio} quantum mechanical calculations \cite{sch} and empirical 
studies \cite{krp1}-\cite{tou} of protein backbone geometry  
have already disclosed that  the backbone bond angle 
$\tau_{NC} \equiv$ (N-C$_\alpha$-C) about the C$_\alpha$ carbons might 
oscillate quite substantially. The range of variations 
can be as large as 8.8$^o$ \cite{krp2}. This corresponds to a shift of $\sim \! 0.6$ \AA ~ in the relative 
positioning of two consecutive C$_\alpha$ carbons. A deviation of this size  from the ideal value 
can be subjected to experimental scrutiny in  X-ray experiments that reach 
sub-{\AA}ngstr\"om resolution. Indeed, on the basis of 
existing data the authors \cite{krp1}-\cite{tou} have already reported  that the deviations in
the values of the $\tau_{NC}$ angle are systematic, and in particular that these deviations
reflect the local  secondary structure.

The $\tau_{NC} $ angles are primarily affected by the backbone.  
As such, their values relate directly to the two standard Ramachandran angles, that form the basis 
for structure validation. As a consequence, the literature \cite{sch}-\cite{tou} has until now mainly
concentrated on the effects that potential deviations of $\tau_{NC} $ from ideality have on
the backbone geometry. Here we extend this analysis to the side-chains:
The fluctuations in the lengths of the covalent bonds in 
the C$_\alpha$ tetrahedron are no more than around 0.1 \AA  ~ which is much
less than the potential $\sim \! 0.6$ \AA ~ shift in the relative positioning of 
two consecutive C$_\alpha$ carbons, due to $\tau_{NC} $ fluctuations  \cite{krp1}-\cite{tou}.  
This proposes that any deviation of $\tau_{NC}$ from its
ideal value inevitably propagates  to the side-chain dependent
$\tau_{N\beta} \equiv$ (N-C$_\alpha$-C$_\beta$)  and  $\tau_{C\beta}\equiv$ (C-C$_\alpha$-C$_\beta$)  
bond angles, and  this  should lead to observable effects in the angular positions of the 
side-chain C$_\beta$ atoms.


In this article we first analyze PDB data to find whether there are experimental  variations in the  tetrahedral 
angles around the C$_\alpha$.
In particular,  we extend the analysis of \cite{krp1}-\cite{tou} to study correlations between 
the side-chain dependent angles $\tau_{N\beta}$  and  $\tau_{C\beta}$   that determine the C$_\beta$ orientations,
and the local secondary structure of the backbone.  Since the side-chain atom positions are not easily described in terms of
the backbone  Ramachandran angles,  we start by developing new visualization tools. 
In line with  \cite{krp1}-\cite{tou} we observe that the local secondary structure has a systematic effect 
on the relative tetrahedral position of the C$_\beta$ carbon. We then proceed to utilize
our visualization tools  to develop theoretical arguments. 
We  propose a coarse-grained framework that computes how the observed 
direction of the C$_\beta$ evolves along the backbone. In particular, we argue that  the 
direction of the C$_\beta$ can be computed  from the soliton solution of a 
discrete nonlinear Schr\"odinger (DNLS) equation. 
The DNLS soliton already shares a remarkable history with protein 
research \cite{dnls}. Both the DNLS equation and its soliton solution
were first introduced by Davydov to describe the propagation of 
energy along $\alpha$-helices \cite{davy}. He also proposed that since the propagation leads to a local 
deformation of the protein shape,  a trapped soliton is a natural cause for the
protein to fold. Here we first argue on general grounds that 
the DNLS soliton solution can  be utilized to determine  the secondary structure dependence in
the relative direction of the C$_\beta$ atoms along
the backbone. We then consider an explicit example to illustrate our general arguments. 
The example we consider is the 35-residue 
subdomain of the villin headpiece  with PDB code 1YRF. It is a paradigm protein that has been studied widely
in theoretical approaches to protein folding.


\section{II: Visualization of the C$_\alpha$ tetrahedron}

We start by visual analysis of crystallographic protein data in PDB. The goal is to reveal any
secondary structure dependence in  the values of $\tau_{NC}$,  and in
the adjacent  
\[
\tau_{N\beta} \  \equiv \  (N-C_\alpha-C_\beta)
\]
and  
\[
\tau_{C\beta} \ \equiv \  (C-C_\alpha-C_\beta)  
\]
bond angles.   
In order to minimize any bias, we inspect several subsets of PDB.
These include the canonical one that comprises all  PDB 
configurations with resolution 2.0 \.A or better, and its two subsets 
with resolution better than 1.5 \.A, and better than 1.0 \.A. We
also inspect  a subset of  the 2.0 \.A  set that contains only those  
proteins that have less than 30$\%$ sequence similarity. 
Our conclusions are independent of the data set, and for 
illustrative purposes we here use the canonical 2.0 \.A set. There are presently
over 30.000 such entries
in PDB.

The Ramachandran angles are defined in terms  of the backbone amide 
planes. As such they are not the most convenient ones
for describing the side-chain geometry. Since both $\tau_{N\beta}$ and $\tau_{C\beta}$ relate to the side-chain 
geometry, we prefer to follow  \cite{dff}  and describe the folded protein structure 
in terms of the geometrically determined backbone discrete Frenet frames (DFF). These frames govern the entire backbone
neighborhood, including the side-chains. But their construction involves {\it only} the C$_\alpha$ coordinates 
${\bf r}_i$ where $i=1,...,N$ label the residues. As such, these frames then 
give a manifestly N, C$_\beta$ and C independent, purely geometric description of the tetrahedral
$sp3$ neighborhood of the C$_\alpha$ atoms. 

The backbone tangent vectors are
\begin{equation}
\mathbf t_i = \frac{ {\bf r}_{i+1} - {\bf r}_i  }{ |  {\bf r}_{i+1} - {\bf r}_i | }
\la{t}
\end{equation}
The unit binormal vectors are
\begin{equation}
\mathbf b_i = \frac{ {\mathbf t}_{i-1} - {\mathbf t}_i  }{  |  {\mathbf t}_{i-1} - {\mathbf t}_i  | }
\la{b}
\end{equation}
The unit normal vectors 
\begin{equation}
\mathbf n_i = \mathbf b_i \times \mathbf t_i
\la{n}
\end{equation}
The orthogonal triplets ($\mathbf n_i, \mathbf b_i , \mathbf t_i$) determine the discrete Frenet frame 
at each of the positions $\mathbf r_i$ of the C$_\alpha$ carbons. 
Note that if the tangent vectors and the distances between the C$_\alpha$ are known, 
we can reconstruct the entire C$_\alpha$ backbone using 
\begin{equation}
\mathbf r_k = \sum_{i=1}^{k-1} |\mathbf r_{i+1} - \mathbf r_i | \cdot \mathbf t_i
\la{dffe}
\end{equation}
For the initial condition we can utilize Galilean invariance to take $\mathbf r_1 = 0$, and $\mathbf t_1$ to 
point into the direction of the positive $z$-axis.  
In particular, (\ref{dffe}) does not involve the vectors $\mathbf n_i$ and $\mathbf b_i$.

We introduce the backbone bond angles
\begin{equation}
\cos \kappa_{i+1} \ \equiv \ \cos\kappa_{i+1 , i} = {\bf t}_{i+1} \cdot {\bf t}_i
\la{bond}
\end{equation}
and the backbone torsion angles
\begin{equation}
\cos \tau_{i+1} \ \equiv \ \cos\tau_{i+1,i} = {\bf b}_{i+1} \cdot {\bf b}_i
\la{tors}
\end{equation}
Note that  both angles are manifestly 
independent of the N, C$_\beta$ and C atoms that are covalently bonded to the
C$_\alpha$ atoms, in this sense they provide an unbiased set of coordinates for describing the 
positions of these atoms. If these angles are known, 
we can use 
\[
\left( \begin{matrix} {\bf n}_{i+1} \\  {\bf b }_{i+1} \\ {\bf t}_{i+1} \end{matrix} \right)
= 
\left( \begin{matrix} \cos\kappa \cos \tau& \cos\kappa \sin\tau & -\sin\kappa \\
-\sin\tau& \cos\tau & 0 \\
\sin\kappa \cos\tau& \sin\kappa \sin\tau & \cos\kappa \end{matrix}\right)_{\hskip -0.1cm i+1 , i}
\left( \begin{matrix} {\bf n}_{i} \\  {\bf b }_{i} \\ {\bf t}_{i} \end{matrix} \right) 
\]
\begin{equation}
 \equiv \ {\mathcal R}_{i+1,i} 
\left( \begin{matrix} {\bf n}_{i} \\  {\bf b }_{i} \\ {\bf t}_{i} \end{matrix} \right)
\la{DFE2}
\end{equation}
to iteratively construct the Frenet frame at position $i+i$ 
from the frame at position $i$. Once we have all the frames, we can 
proceed to construct  the entire backbone using
(\ref{dffe}).

The bond and torsion angles have a natural interpretation in terms of  
the canonical latitude and longitude angles of a two-sphere $\mathbb S^2$. 
In the sequel we find it useful to extend the range of $\kappa_i$ to $[-\pi, \pi] \ mod(2\pi)$.
But we introduce no change in the range of $\tau_i \in [-\pi, \pi] \ mod(2\pi)$. 
We compensate for this two-fold covering of $\mathbb S^2$  by  the following discrete $\mathbb Z_2$ 
symmetry
\begin{equation}
\begin{matrix}
\kappa_k \ \to \ - \kappa_k \ \ \ {\rm for ~ all~} \ k \geq i \\
\hskip -1.9cm \tau_i \ \to \ \tau_i - \pi
\end{matrix}
\label{Z2}
\end{equation}
It inverts the directions of the vectors $\mathbf n_i$ and $\mathbf b_i$ but has no effect on the $\mathbf t_i$ and consequently
leaves the backbone intact. For details we refer to \cite{dff}.

We use the discrete Frenet  frames to display each atom
in the way, how the atom is seen on the surface of a sphere that surrounds 
an imaginary observer who roller-coasts  the backbone along the 
C$_\alpha$ atoms, so that the gaze direction is always fixed towards the next C$_\alpha$ and with local orientation
determined by the DFF frames \cite{dff}.

In Figure 1 we show the statistical angular distribution of the backbone N and C atoms, and in Figure 2
we show the same for the 
side-chain C$_\beta$ atoms in our PDB data set  as seen by the Frenet frame observer 
who moves through all the proteins in our data set.
The sphere is  centered at the C$_\alpha$, and its radius coincides with
the length of the (approximatively constant) covalent bond. We take the vector
$\bf t$ that points towards the next C$_\alpha$ to be  in the direction of the positive $z$-axis, towards the north-pole of the
sphere. With
$\mathbf n$ in the direction of positive $x$-axis we have  a right-handed Cartesian coordinate system. 
We introduce the canonical spherical coordinates ($\theta,\varphi$) 
to describe the distributions. The angle $\theta \in [0,\pi]$ measures 
latitude from the positive $z$-axis, hence it describes the distribution of the bond angles $\kappa_i$. 
The angle $\varphi \in [-\pi,\pi]$ measures longitude  in a counterclockwise direction from the $x$-axis {\it i.e.} from the
direction of $\mathbf n$ towards that of $\mathbf b$, with $\varphi=0$ at the $x$-axis. Consequently it describes the distribution
of the torsion angles $\tau_i$.
%
\begin{figure}[h]
        \centering
                \includegraphics[width=0.45\textwidth]{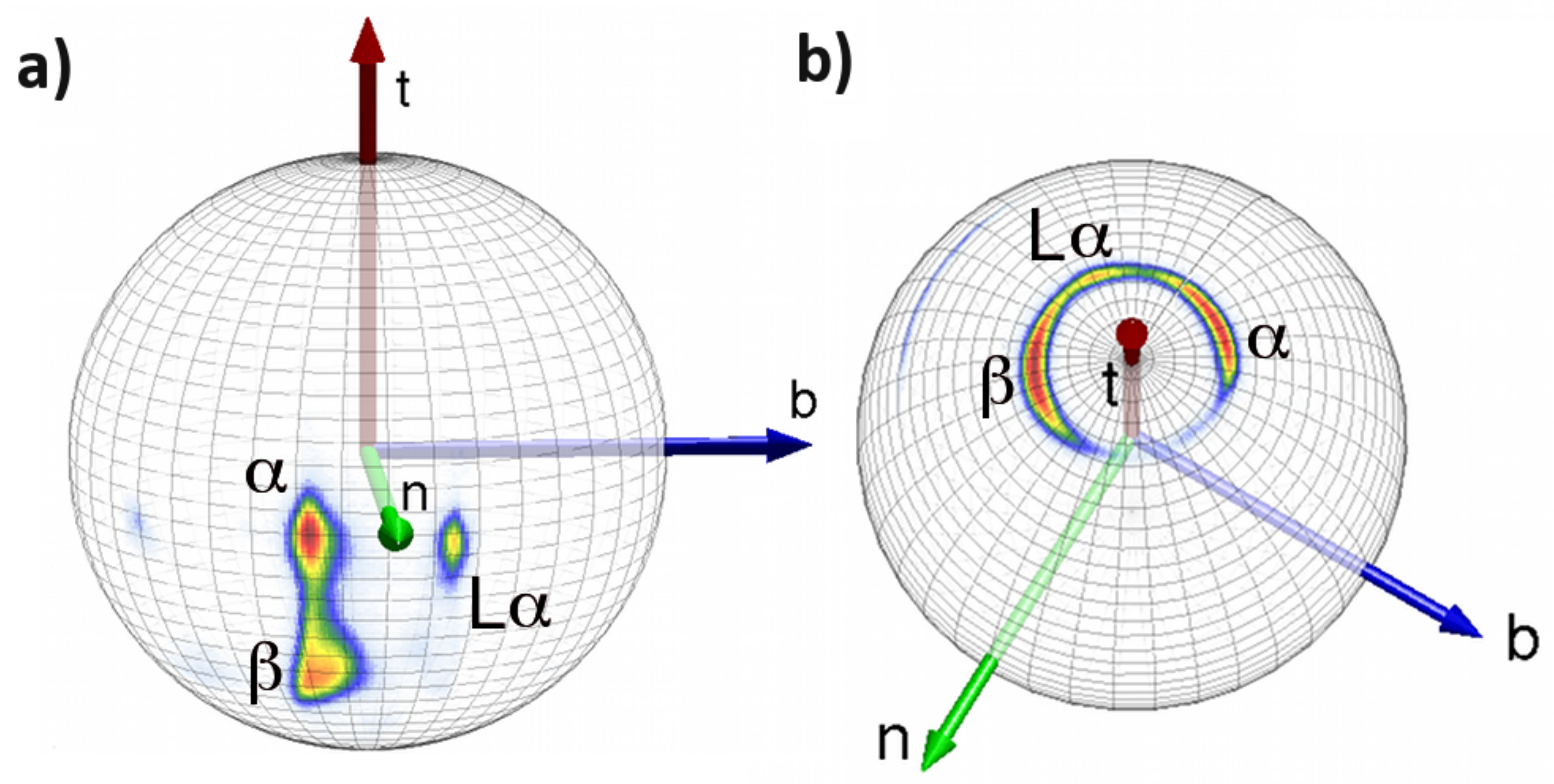}
        \caption{{ 
     (Color online) The directions  of a)  backbone N-atoms and  b)  backbone C-atoms 
     as seen by a Frenet frame observer  located at  the C$_\alpha$ carbon 
     which is at the center  of the sphere.     
     In a) the smaller, more point-like  direction of backbone N atoms corresponds to the  {\tt L}-$\alpha$  Ramachandran region.
     The larger  region forms a segment of the great circle $\varphi \approx -15^{\mathrm o}$. Loops interpolate latitudinally 
     between $\alpha$-helices and $\beta$-sheets.  
      In b) the directions of backbone C form a segment 
      of a small circle around $z$-axis, with $\theta \approx  20^{\mathrm o}$.
       }}
       \label{Figure 1}
\end{figure}

\begin{figure}[h]
        \centering
                \includegraphics[width=0.475\textwidth]{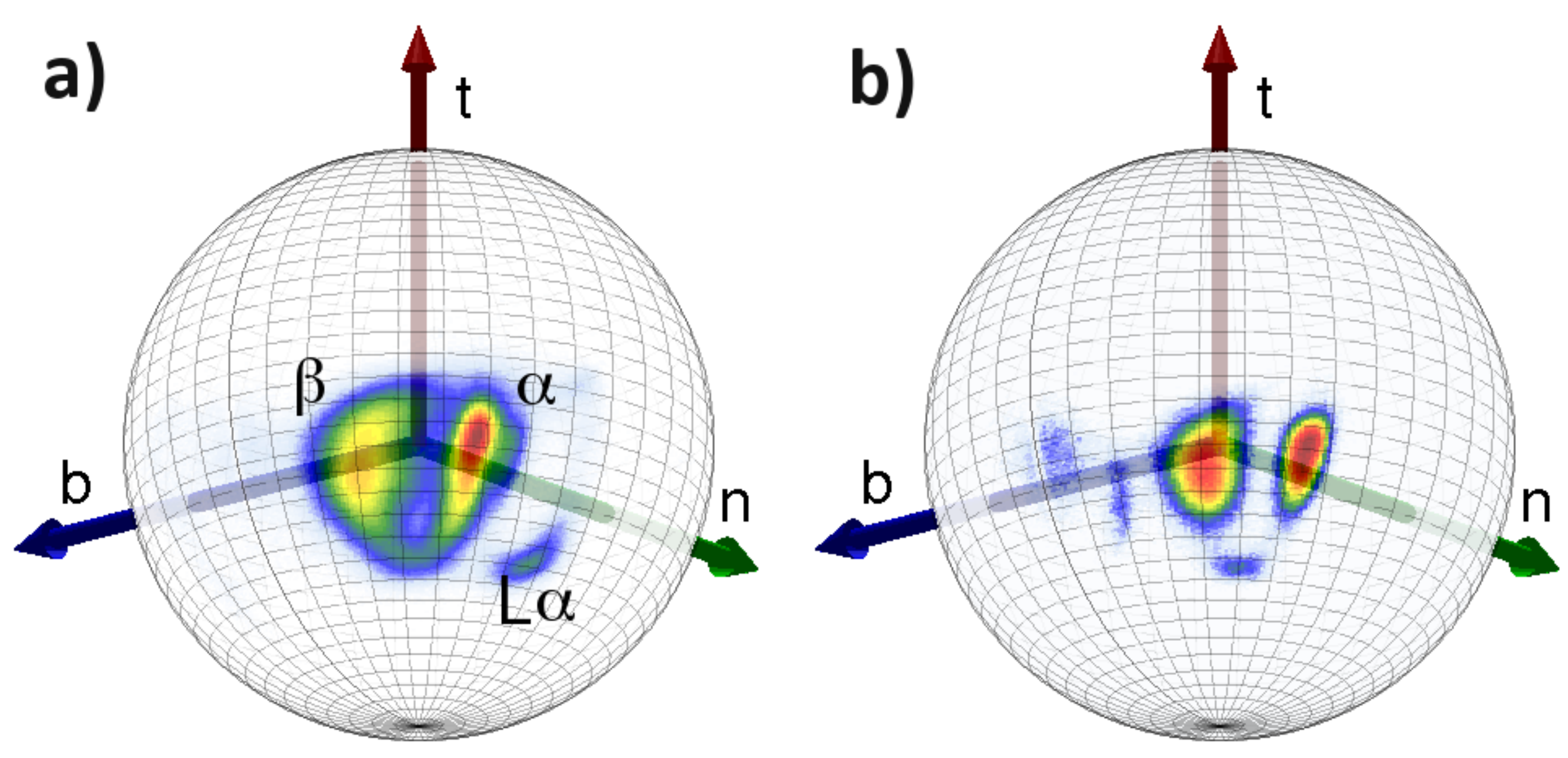}
        \caption{{ 
     (Color online)  The distribution of the C$_\beta$ directions in the Frenet frame. In  a) we have  
     all amino acids (including proline but excluding glycine     
     that has no C$_\beta$).  
     In b) we show only proline.  Comparison between a) and b) exemplifies how the C$_\beta$ direction can depend on the   individual  amino acid.
      We have  chosen
     proline in b) as it  is particularly interesting due to the way how it appears in Figure 3b). 
}}
       \label{Figure 2}
\end{figure}

We find that in the Frenet frame coordinate system, 
the N and  C  oscillations shown in Figures 1a) and 1b)  
are fully separated into the  locally orthogonal $\theta$ and $\varphi$ directions, respectively. This would 
certainly not be the case in a generic coordinate system.  Furthermore, 
secondary structures such as  $\alpha$-helices, $\beta$-sheets, loops and 
left-handed $\alpha$-regions are all clearly identifiable in Figures 1.
Figure 2a) reveals how the  N and C oscillations  of Figure 1, through  the covalent bonds that form the $sp3$ tetrahedron around C$_\alpha$,  combine into a horseshoe (annulus) shaped 
nutation of C$_\beta$. As visible in the Figure, this  nutation reflects  the local secondary structure environment 
in  an equally  systematic manner as Figures 1. 

%
\begin{figure}[h]
        \centering
                \includegraphics[width=0.5\textwidth]{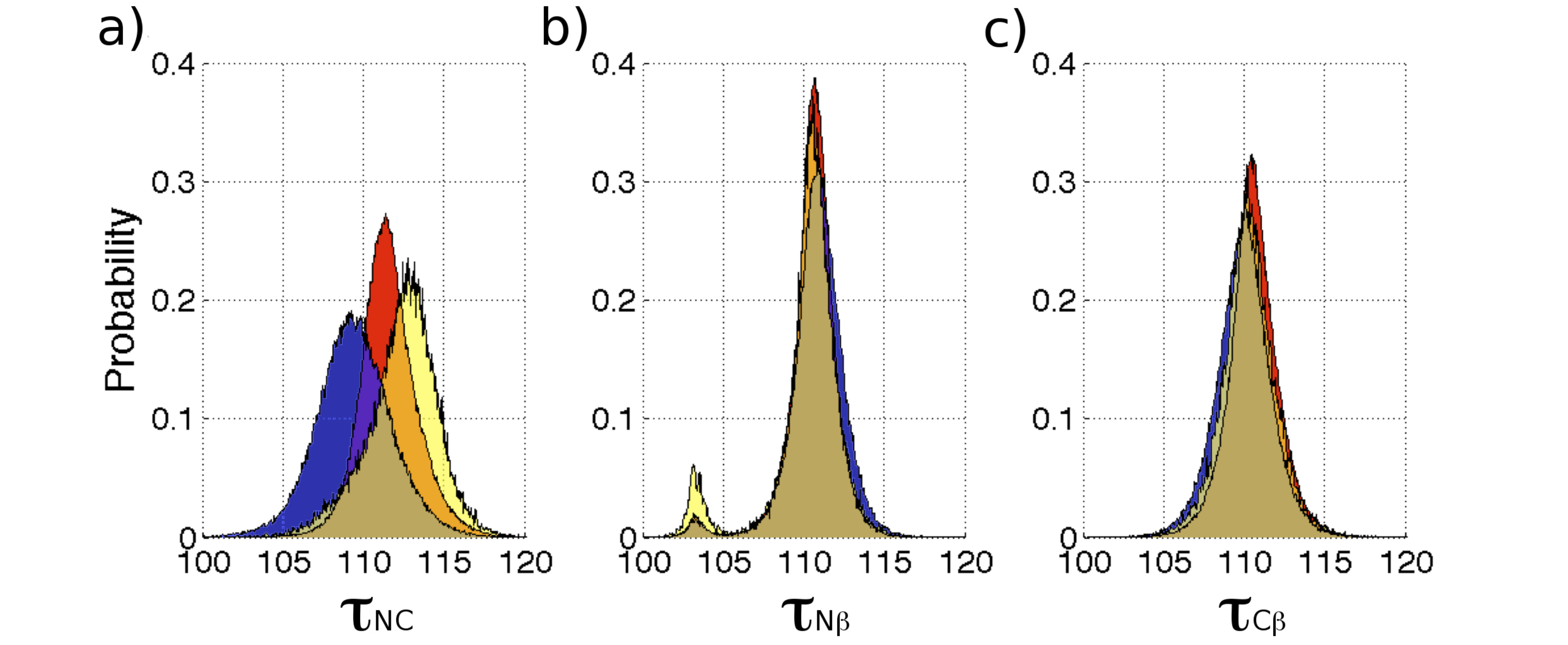}
        \caption{{ (Color online)
     The normalized probability density angular distribution 
     of the a) $\tau_{NC}$,  b)  $\tau_{N\beta}$ and c) $\tau_{C\beta}$
     angles in degrees, with $\alpha$-helices in red (grey), $\beta$-strands in blue (dark grey),  and 3/10 helices in yellow. 
      The secondary peak in b) is due to {\it cis}-peptide prolines.  Only in Figure a) are the different  secondary structures
      visibly separated from each other. In Figures b) and c) they are practically fully overlapping, the only notable effect is 
      the somewhat higher propensity of 3/10 helices in connection of {\it cis}-peptide prolines.  }}
       \label{Figure 2}
\end{figure}

The pattern of angular separation in the N and C oscillations in Figure 1 reveals that the
underlying ideal tetrahedral covalent symmetry around C$_\alpha$ is not transferable along the protein backbone
\cite{krp1}-\cite{tou}. 

We
proceed to Figures 3 a)-c) where we plot the tetrahedral bond angles $\tau_{NC}$,  
$\tau_{C\beta}$ and  $\tau_{N\beta}$  jointly for  the $\alpha$-helices, $\beta$-strands and $3/10$-helices. 
As in figure 1 the loops interpolate continuously between these regular secondary structures. 
The Figure 3a) clearly reveals that  at the level of the $\tau_{NC}$ angle the transferability of the
tetrahedral symmetry is absent
in a systematic and secondary structure dependent manner. But neither $\tau_{C\beta}$  nor $\tau_{N\beta}$
show any sign whatsoever of secondary structure dependence. The distributions are practically the 
same, independently of the secondary structure. (The isolated small peak in Figure  3b)   is due to  
proline  {\it cis}-peptide groups.) 

As such, the distribution in both Figures 3b) and 3c)  is 
what we would expect in the case of ideal, transferable $sp3$ tetrahedral symmetry. 
In each Figure the average value is around $111^{\rm o}$ and there are secondary 
structure independent fluctuations that are in line with quantum mechanical
estimates and \cite{sch} and empirical studies \cite{krp1}-\cite{tou}.
But the fact that  $\tau_{NC}$  in Figure 3a) displays  clearly visible and
systematic secondary structure dependence 
makes it plain and clear that the paradigm of transferable
tetrahedral covalent symmetry around the  C$_\alpha$ carbon  is absent. 
Furthermore, the way how transferability becomes violated
reflects the wider secondary structure environment
of the amino acid along the protein backbone.

Since the deviation from the ideal tetrahedral symmetry is 
organized  in the same way how the proteins are folded,
these two must share a common origin.  
But we do not have any physical explanation why the lack of ideal symmetry  is only
visible in  $\tau_{NC}$. We suspect this has to do with existing experimental refinement methods,
the way how refinement tension is distributed between the backbone and side-chains.
The high resolution crystallographic data which becomes available in future third and fourth generation
experiments should help to clarify this.


%
%
%
%
%
%
%
%
%
%
%
%
%
%
%
%
%
%
%

\section{III: Solitons and  side-chains}

Any molecular dynamics approach to protein folding that we are familiar with,
utilizes the harmonic approximation 
(\ref{ba}) for bond and torsion angles. Here $\kappa_0$  is in general amino acid dependent but secondary structure 
independent equilibrium value of the bond angle.  
But from Figure 3a) we observe, that for the $\tau_{NC}$ angle 
there are three different major equilibrium values. These equilibrium bond angle values are 
amino acid independent, but do depend in a nontrivial manner on the local secondary structure:
The three peaks in Figure 3a)
correspond to the $\alpha$-helix, $\beta$-strand, and $3/10$-helix while for a loop the values 
of the corresponding equilibrium $\kappa_0$ interpolates between these three ground state values. Since each of these
secondary structures are characterized by a different equilibrium 
value of bond angle $\kappa$, to the leading order we may 
take $\kappa_0$ to be a function of $\kappa$. By expanding to leading order we get
\[
\kappa_0  \to \kappa_0(\kappa) \ \approx \ \kappa^{(0)}_0 + \kappa^{(1)}_0 \cdot \kappa + \kappa^{(2)}_0 \cdot \kappa^2
+ \mathcal O (\kappa^3)
\]
where the $\kappa^{(i)}$ are independent of the local value of $\kappa$. The first two terms simply 
renormalize the values of $\omega_\kappa$ and $\kappa_0$ in
(\ref{ba}). But the third term is conceptually different. It 
introduces an anharmonic correction. We conclude that after redefinitions 
of the parameters, in the leading order the potential obtains the functional form
\begin{equation}
E_{bond} \ \sim \ q\cdot (\kappa^2 - m^2)^2
\label{ba2}
\end{equation}
We argue that based on the Figure 3a), the anharmonic corrections are already visible in the existing 
high resolution X-ray data.  In order to answer the theoretical challenge that this poses,  
we propose  to improve existing MD force fields, to account for 
the anharmonic corrections in the bond angle contribution. 

\subsection{A: Backbone energy}

The presence of an anharmonic correction in the bond angle energy 
has important implications to the way how proteins
fold.  For this we  start with a simple example. We consider
the anharmonic potential (\ref{ba2}) in the presence of a 
single coordinate  $x$ on a line. The Newton's equation is 
\[
m \ddot x =  - \frac{dV}{dx} 
\]
We take the potential to have the form
\begin{equation}
V(x) =  \frac{\sigma}{4} \hskip 0.3mm  \, (  x+b)^2 \cdot  (x-a)^2
\label{Vx}
\end{equation}
We introduce
\[
c =  -\frac{1}{2} (b+a)
\]
and define 
\[
y  = x -  \frac{1}{2}(a-b)
\]
to arrive at the equation of motion
\[
\ddot y  =  - \frac{\sigma}{m} { y} ({y}^2 - c^2)
\]
which is essentially the continuum nonlinear Schr\"odinger equation (NLSE) \cite{dnls}, \cite{davy}
Note that the potential has the symmetric form (\ref{ba2}). This equation is solved by
\[
{ y }(t) \ = \  c \cdot \tanh [\, c \, \sqrt \frac{ \sigma}{2m}  (t-t_0) ]  
\]
 \[
\Rightarrow \  x(t) \ = \  y(t) + \frac{1}{2} (a-b) 
\]
\begin{equation}
= \ - \, \frac{ b \cdot  e^{c \, \sqrt \frac{ \sigma}{2m}  (t-t_0) } \ -  \  a  \cdot
e^{- c \, \sqrt \frac{ \sigma}{2m}  (t-t_0) } } {  \cosh [ c \, \sqrt \frac{ \sigma}{2m}  (t-t_0) ] }
\label{dark}
\end{equation}
This is the hallmark NLSE soliton  configuration, so called dark soliton solution of the NLSE equation.  It 
interpolates between the two uniform ground 
states at $x = a $ and $x=-b$ when $t \to \pm \infty$. 
The parameters $a$, $b$, $t_0$ and the 
combination 
\[
c \, \sqrt \frac{ \sigma}{2m}
\] 
are the canonical ones that characterize the asymptotic values
of $x(t)$ {\it i.e.} minima of the potential,  and the size and location of the soliton.

For finite $t$ the soliton (\ref{dark}) describes  a configuration with an energy
above  the uniform ground state $x\equiv a$  (or $x \equiv b$). Nevertheless, it  
can not decay into  $x  \equiv a$ (or $x \equiv b$)
through any kind of continuous finite energy 
transformation.  In particular, a soliton configuration such as (\ref{dark}) 
can not be obtained from any approach that only accounts for perturbations that
describe small localized  fluctuations around the uniform background ground state.  

%

%
%
%
%
%
%
%
%
%
%
%
%
%
%
%
%
%
%
%
%
%
%
In \cite{oma2}-\cite{and1}, it has been shown that  the soliton profile (\ref{dark}) can be used to
describe  loops 
in folded proteins.  For this one merges general geometric arguments with 
the concept of universality \cite{widom}-\cite{fisher}, to arrive at the following
simplified, coarse-grained energy function for the backbone bond and torsion angles \cite{oma1}, \cite{ulf},
\[
E = - \sum\limits_{i=1}^{N-1}  2\, \kappa_{i+1} \kappa_{i}  + \sum\limits_{i=1}^N
\biggl\{  2 \kappa_{i}^2 + q\cdot (\kappa_{i}^2 - m^2)^2  
\]
\begin{equation}
 \left. + \frac{d_\tau}{2} \, \kappa_{i}^2 \tau_{i}^2   - {b_\tau}\kappa_i^2 \tau_i
- a_\tau  \tau_{i}   +  \frac{c_\tau}{2}  \tau^2_{i} 
\right\} 
\la{E1}
\end{equation}
where $\kappa_i$ and $\tau_i$ are the backbone bond and torsion angles (\ref{bond}), (\ref{tors}).
Unlike force fields in molecular dynamics, the
energy function (\ref{E1}) does not purport to explain the fine 
details of the atomic level mechanisms that give rise to protein folding. Instead, in line with general
principles of effective Landau-Lifschitz  theories it describes the properties of a folded protein 
backbone in terms of universal physical arguments. Indeed, according to the concept of universality
\cite{widom}-\cite{fisher} the energy function
(\ref{E1})  can be viewed as the universal  long distance limit that emerges from any atomic level 
energy function  when the internal energy is coarse-grained
to include only the backbone bond and torsion angles. 

In order to construct the soliton solution, we start by introducing the $\tau$-equation of motion 
\[ 
\frac{\partial E}{\partial \tau_i} =  d_\tau \kappa_i^2 \tau_i - b_\tau \kappa_i^2 - a_\tau + c_\tau \tau_i = 0 
\]
\begin{equation}
\Rightarrow \ \tau_{i} [\kappa] = \frac{a_\tau + b_\tau \kappa_i^2}{c_\tau + d_\tau \kappa^2_{i} } 
\la{tauk} 
\end{equation}
Notice that even though there are four parameters in (\ref{tauk}) one of them, the overall scale, drops out.
We then use (\ref{tauk}) to eliminate the torsion angles,  so that the energy for the bond angles becomes
\begin{equation}
E[\kappa] = - \sum\limits_{i=1}^{N-1}  2\, \kappa_{i+1} \kappa_{i}  + \sum\limits_{i=1}^N
 2 \kappa_{i}^2 + V[\kappa_i]  
\la{Ekappa}
\end{equation}
where
\[
V[\kappa]  =  - \left( \frac{b_\tau c_\tau - a_\tau d_\tau}{d_\tau}  \right) \cdot \frac{1}{c_\tau+d_\tau\kappa^2} 
\]
\begin{equation}
- \left( \frac{b_\tau^2 + 8 q m^2}{2b_\tau} \right)
\cdot \kappa^2 + q\cdot \kappa^4
\la{Vkappa}
\end{equation}
Because the first term contains $\kappa$ in the denominator, its variation with $\kappa$  
is not that pronounced as the variation of the other two terms, which are proportional to the second and the fourth power of 
$\kappa$, respectively. Moreover, because $|\kappa|>1$ radian for proteins, it turns out that the first term is 
small in value compared to the other terms.
The second and third terms have then the functional form  of the double well potential 
(\ref{ba2}). 

In applications to folded  proteins the parameters values are such,   
that in the energy ground state both $\kappa$ and $\tau$  acquire a 
non-vanishing value.  In particular, since the functional form of (\ref{Vkappa}) is similar to (\ref{ba2}), (\ref{Vx})
we can expect that there are soliton solutions:

Geometrically, a uniform constant value of the bond and torsion 
angles describes regular protein secondary 
structures. For example,  the standard $\alpha$-helix is
\begin{equation}
\alpha-{\rm helix:} \ \ \ \ \left\{ \begin{matrix} \kappa \approx \frac{\pi}{2}  \\ \tau \approx 1\end{matrix} \right.
\la{bc1}
\end{equation}
and for the standard $\beta$-strand we have
\begin{equation}
\beta-{\rm strand:} \ \ \ \ \left\{ \begin{matrix} \kappa \approx 1 \\ \tau \approx \pi \end{matrix}  \right.
\la{bc2}
\end{equation}
The additional regular secondary structures including 3/10 helices, left-handed helices {\it etc.}  are described
similarly. 

But in addition of constant value configurations, as in (\ref{Vx})  there are also soliton solutions.  
In particular, since protein loops are structures that interpolate between 
different constant values such as (\ref{bc1}), (\ref{bc2}), this means that loops 
correspond to these soliton solutions 
\cite{oma2}-\cite{and1}. In order to construct the relevant soliton, we introduce 
the generalized discrete nonlinear Schr\"odinger (DNLS) equation that derives
from the energy (\ref{E1}).  Variation of this energy {\it w.r.t.} $\kappa_{i}$ and substitution 
of (\ref{tauk}) gives 
\begin{equation}
\kappa_{i+1} = 2\kappa_i - \kappa_{i-1} + \frac{ d V[\kappa]}{d\kappa_i^2} \kappa_i  \ \ \ \ \ (i=1,...,N)
\la{nlse}
\end{equation}
where $\kappa_0 = \kappa_{N+1}=0$.  The exact  soliton solution to the present discrete nonlinear 
Schr\"odinger equation is not known in a closed form. But numerical approximations can be easily constructed
using the procedure described in \cite{nora}. Furthermore, whenever the first term in (\ref{Vkappa}) is small as it is in the
case of proteins, an excellent approximation \cite{and1}
is obtained from the {\it naive} discretization of the 
continuum soliton  (\ref{dark}),
\begin{equation}
\kappa_i  =  \frac{ 
\mu_{1}   \cdot e^{ \sigma_{1} ( i-s)  } - \mu_{2}  \cdot e^{ - \sigma_{2} ( i-s)}  }
{e^{ \sigma_{1} ( i-s) } +  e^{ - \sigma_{2} ( i-s)}   }
\la{bond2}
\end{equation}
Here $s$ is a parameter that determines the backbone site location of the center of the fundamental loop 
that is described by 
the soliton.  The $\mu_{1,2} \in [0,\pi] $ are parameters, their values 
are entirely determined by the adjacent helices and strands: 
Away from the soliton center we  have
\[
\kappa_i \ \to \left\{ \begin{matrix}  \mu_1 \  &  \ \ \ \ i > s \\  -\mu_2 \  &  \ \ \ \ i < s\end{matrix} \right.
\]
and for $\alpha$-helices and $\beta$-strands the $\mu_{1,2}$ values are determined by (\ref{bc1}), (\ref{bc2}).
We remind that negative values of $\kappa_i$ are related to the positive values by (\ref{Z2}).
Note that for $\mu_1 = \mu_2$  and $\sigma_1 = \sigma_2$ we  recover the hyperbolic tangent. In this
case the two regular secondary structures before and after the loop are the same.
Moreover,  {\it only}  the (positive)  $\sigma_1$ and  $\sigma_2$ are 
intrinsically loop specific parameters, they specify the length of the loop and as in the case of the $\mu_{1,2}$, 
they are combinations of the parameters in (\ref{E1}).

Similarly, in the case of the torsion angle there is only one loop specific parameter in (\ref{tauk}): The overall, common
scale of the four parameters is irrelevant in (\ref{tauk}), and two of the remaining three parameters characterize  the regular
secondary structures that are adjacent to the loop, as in (\ref{bc1}), (\ref{bc2}). 

Entire protein loops can be constructed by combining together solitons  (\ref{nlse}), (\ref{tauk}). In \cite{peng} it has been shown 
using the Ansatz (\ref{bond2}) that over 92$\%$ of crystallographic 
PDB configurations can be constructed in terms of 200 explicit soliton profiles. The solitons of the DNLS equation can thus
be interpreted as the modular building blocks of folded proteins.

\subsection{B: Side-chain energy }

%
%
%
%
%
%
%
%
%
%
%
%
%
%
%
%
%
%
%
%
%
%
%
%
%
%
%
%
%
%

We proceed to extend the energy function (\ref{E1}) so that it models the deviations from the paradigm
tetrahedral symmetry 
around the C$_\alpha$ atoms:  The Figures 1a) and 1b) reveal that the directions 
of the backbone N and C atoms oscillate 
in the latitudinal  ($\theta$) and longitudinal ($\varphi$) directions respectively, on the surface of the sphere that 
surrounds the corresponding C$_\alpha$ atom. 
The covalent bond structure that forms the $sp3$ tetrahedron of the C$_\alpha$ atom combines  these two oscillations 
into the annulus (horseshoe) shaped C$_\beta$ nutation of Figure 2a). Consequently 
the natural dynamical variable that describes the nutation of C$_\beta$  on the surface of the sphere 
is the canonically parametrized 
three component unit vector 
\begin{equation}
\mathbf u = \left( \begin{matrix} \sin\theta \cdot \cos \varphi \\ \sin\theta \cdot \sin \varphi \\ \cos \theta \end{matrix} \right) 
\la{uvec}
\end{equation}
In order to account for the C$_\beta$ nutation contribution to the protein free energy, we then
need to augment (\ref{E1}) by terms that engage the  
additional variables  ($\theta_i, \varphi_i)$. 

The latitude angle $\theta_i$ is counted from the direction of the corresponding  Frenet frame
tangent vector $\mathbf t_i$.
Consequently it remains  invariant under the rotations of the local Frenet frames  around the direction of $\mathbf t_i$
\cite{dff}. Thus it can only couple to other frame rotation invariant
quantities. There are two natural terms,
\[
|\mathbf  t_i \times \mathbf  u_i| = \sin \theta_i 
\]
and 
\[
\mathbf t_i \cdot \mathbf u_i = \cos \theta_i 
\]
In the leading order we only account for local interactions. When we also demand 
invariance under the $\mathbb Z_2$ gauge transformation (\ref{Z2}) 
we conclude that to the leading order 
the corresponding free energy contribution should have the form
\begin{equation}
E_\theta  =
\sum\limits_{i=1}^N  
f_i(\kappa_i^2)
\,   |\mathbf t_i \times \mathbf  u_i|  + g_i(\kappa_i^2)  \, \mathbf  t_i \cdot \mathbf u_i   + \dots 
\la{E2a1}
\end{equation}
According to Figure 2a) the range of variations in $\theta_i$ are quite small and we estimate
that the center of the annulus-like region is near 
\[
<\theta> \ \approx \ 113.4^{\rm o}  
\]
We Taylor expand (\ref{uvec}) around this value so that we have to the leading
order
\begin{equation}
E_\theta  =
\sum\limits_{i=1}^N
\left\{  \frac{d_\theta}{2} \, \kappa_i^2 \theta_i^2  -  b_\theta \kappa_i^2  
\theta_i - a_\theta \theta_i   +  \frac{c_\theta}{2}  \theta^2_i 
\right\}   + \dots
\label{E2a}
\end{equation}
The ensuing equation of motion is
\begin{equation}
\theta_i = \frac{a_\theta + b_\theta \kappa_i^2}{c_\theta + d_\theta \kappa_i^2} 
\la{thei}
\end{equation}
As in the case of (\ref{tauk}) we conclude that the overall scale of the parameters drops out and this leaves us
with three independent parameters. In the case of a short loop that we can model in terms of a single soliton 
like (\ref{bond2}), two of the parameters  become determined by the value of $\theta_i$ 
in the regular secondary structures that are adjacent to the loop.  This leaves us
with only one loop specific parameter. 

The longitude $\varphi_i$ in (\ref{uvec}) is measured from the direction of the Frenet 
frame normal vector $\mathbf n_i$. Under the local rotations of the Frenet frames 
\[
\left( \begin{matrix} \mathbf n_i \\ \mathbf b_i \end{matrix} \right) \ \to \ 
\left( \begin{matrix} \cos\Delta_i & \sin\Delta_i \\ -\sin\Delta_i & \cos \Delta_i  \end{matrix} \right)
\left( \begin{matrix} \mathbf n_i \\ \mathbf b_i \end{matrix} \right) 
\]
around the  
tangent vectors $\mathbf t_i$ by an angle $\Delta_i$ we then have
\[
\varphi_i \  \to \ \varphi_i + \Delta_i
\]
Thus we may couple $\varphi$ to the torsion angle as follows,
\[
\varphi_i + \sum\limits_{k=1}^i \tau_k
\]
This  combination is invariant under the local rotations of the Frenet frame around $\mathbf t_i$.
Since the $\tau_k$ depend on the backbone angles according to (\ref{tauk}), we can again 
Taylor expand the ensuing energy contribution. 
From Figure 2a) we estimate that for the center of the annulus 
\[
<\varphi> \ \approx \ 139.5^{(\rm o)}
\]
Following (\ref{E2a})  we then Taylor expand the $\varphi$ contribution to free energy around this value
to conclude that to the leading order  we have (in Frenet frames)
\begin{equation}
E_\varphi  =  \sum\limits_{i=1}^N
\left\{  \frac{d_\varphi}{2} \, \kappa_i^2 \varphi_i^2  -  b_\varphi \kappa_i^2  \varphi_i - 
a_\varphi \varphi_i   +  \frac{c_\varphi}{2}  \varphi^2_i 
\right\} +\dots
\label{E2b}
\end{equation}
The equation of motion has the same functional form with (\ref{tauk}), (\ref{thei})
\begin{equation}
\varphi_i = \frac{a_\varphi + b_\varphi \kappa_i^2}{c_\varphi + d_\varphi \kappa_i^2}
\la{phii}
\end{equation}
Again only three of the four parameters in $\varphi$ are independent, the 
overall scale drops out. 

We confirm that the functional forms (\ref{E2a}) and (\ref{E2b}) are in line with the annulus-like (horseshoe-like)
form of the C$_\beta$ nutation in Figure 2a). For this 
we stereographically project the C$_\beta$ distribution in
Figure 2a) onto the complex plane. Despite the nonlinear nature of the standard 
stereographic projection the annulus-like shape is more or less retained. Let  $r$ be the  approximate radius of 
a thin annulus on the complex plane and let $(\theta_0, \varphi_0)$ be the location
of its center. In the limit where the corrections to the round circular profile of the thin annulus become small 
we can determine the approximate form of the C$_\beta$ nutation region from
\begin{equation}
( \tan \theta e^{i \varphi} - \tan \theta_0 e^{i \varphi_0} ) ( \tan \theta e^{-i \varphi} - \tan \theta_0 e^{-i \varphi_0} ) = r^2
\label{zz}
\end{equation}
Note that this is invariant under local frame rotations. We  re-write  $\theta$ in (\ref{thei}) as follows, 
\[
\theta = \theta_0 + \frac{ 1 } { c + d \kappa^2 } 
\]
We substitute this into (\ref{zz}) and Taylor expand to find that  
to {\it leading order} it makes sense to
parametrize $\varphi$ by an expression of the functional form (\ref{phii}).

We make the following remark: 
When we combine (\ref{E1}), (\ref{E2a}) and (\ref{E2b}) we arrive at the total energy
\begin{equation}
E = - \sum\limits_{i=1}^{N-1}  2\, \kappa_{i+1} \kappa_{i}  + \sum\limits_{i=1}^N
\biggl\{  2 \kappa_{i}^2 + q\cdot (\kappa_{i}^2 - m^2)^2  \biggr\}
\la{EA}
\end{equation}
\begin{equation}
+ \sum\limits_{i=1}^N \biggl\{ \frac{d_\tau}{2} \, \kappa_{i}^2 \tau_{i}^2   - {b_\tau}\kappa_i^2 \tau_i
- a_\tau  \tau_{i}   +  \frac{c_\tau}{2}  \tau^2_{i} \biggr\}
\la{EB}
\end{equation}
\begin{equation}
+ \sum\limits_{i=1}^N
\left\{  \frac{d_\theta}{2} \, \kappa_i^2 \theta_i^2  -  b_\theta \kappa_i^2  
\theta_i - a_\theta \theta_i   +  \frac{c_\theta}{2}  \theta^2_i 
\right\}   + \dots
\la{EC}
\end{equation}
\begin{equation}
+ \sum\limits_{i=1}^N
\left\{  \frac{d_\varphi}{2} \, \kappa_i^2 \varphi_i^2  -  b_\varphi \kappa_i^2  \varphi_i - 
a_\varphi \varphi_i   +  \frac{c_\varphi}{2}  \varphi^2_i 
\right\} 
\la{ED}
\end{equation}
We have already established that protein backbones can be described in terms of soliton solutions to (\ref{EA}),
(\ref{EB}).  According to (\ref{tauk}), (\ref{thei}), (\ref{phii}) the presence of (\ref{EC}) and (\ref{ED}) does not
change the functional form of the effective $\kappa_i$ energy (\ref{Ekappa}), (\ref{Vkappa}), all three variables ($\tau_i, 
\theta_i, \varphi_i$) are similarly slaved to the bond angles $\kappa_i$. In particular, from Figure 2a) we conclude that the contribution of (\ref{EC}) and (\ref{ED}) to the full energy must be minuscule:
The range of  variations in the variables  ($\theta_i, \varphi_i$) 
is relatively small.  (This is {\it not} the case with $\tau_i$, see for example Figure 4 below.) 
Thus  the values of (\ref{EC}) and (\ref{ED}) 
show very little variation, and in comparison to (\ref{EB}) these two terms can be treated as if they were tiny perturbations. 

Indeed, in the case of proteins the two terms  (\ref{EC}) and (\ref{ED})
make no contribution to the total energy that we are able to observe.
The variables  ($\theta_i, \varphi_i$) are entirely slaved
by the DNLS soliton profile of the backbone bond angles  $\kappa_i$. Since the direction of the 
vector (\ref{uvec}) that specifies the position of the C$_\beta$ carbon is slaved  to $\kappa_i$, the deviation from 
the ideal tetrahedral symmetry in the C$_\alpha$ covalent bond geometry is determined by the local
secondary structure environment of the amino acid.

%
%
%
%
%
%
%
%
%
%
%
%
%
%
%
%
%
%
%
%
%

\subsection{C: Comment on parameters}

The energy function (\ref{EA})-(\ref{ED}) introduces eleven essential parameters, when we account for the 
overall scales in (\ref{tauk}), (\ref{thei}), (\ref{phii}). According to \cite{peng},  no more than 200 different parameter
sets are needed to describe over 92$\%$ of  high resolution structures in PDB with a precision of 
around 0.6 \AA ~ in RMSD for the C$_\alpha$. The solitons are like 
modular components from which the folded proteins are built. 
At the moment we do not have a method to compute the parameters directly from the sequence. However, even in
its present form the approach can be subjected to a stringent experimental scrutiny: A typical super-secondary structure
described by a soliton such as a helix-loop-helix consists of around 15 amino acids. If we assume that the bond lengths
are fixed, this leaves us with 60 unknown coordinates for the C$_\alpha$ and C$_\beta$ atoms. Since there are only 11 
essential parameters in (\ref{EA})-(\ref{ED}), 
we have a highly under-determined set of equations. Consequently the model is predictive, 
a comparison with experimental
structures is directly testing  the physical principles on which (\ref{EA})-(\ref{ED}) is based, {\it even though} we 
are not yet able to compute the parameters from the sequence.

%
%
%
%
%
%
%
%
%
%
%
%
%
%
%
%
%
%
%
%
%

\section{IV: Example: Villin headpiece HP35}

As an example we consider the chicken villin headpiece subdomain HP35. We use the x-ray structure with
PDB code 1YRF. The HP35  is a naturally existing 35-residue 
protein with three $\alpha$-helices separated from each other by two loops. It continues to  be  the subject 
of very extensive studies both experimentally \cite{meng}-\cite{wic} and {\it in silico} \cite{pande2}-\cite{sha}, and  
\cite{sha} reports on a molecular dynamics construction with overall backbone RMSD 
accuracy around one {\AA}ngstr\"om. 

In Figure 4 we have the ($\kappa_i, \tau_i$) spectrum that we compute from the PDB data of 1YRF.
In the Figure 4a) we use the standard convention that bond angles take values
in the range $[0,\pi]$. In the Figure 4b) we have extended the range to $[-\pi,\pi]$.
This introduced the $\mathbb Z_2$ gauge transformation structure (\ref{Z2}). In Figure 4b) we
have applied the gauge transformation to disclose the solitons. We clearly have two solitons
with the DNLS profile (\ref{bond2}),  separated from each other by regions
with $\kappa \approx \pm 1.57 $ and $\tau \approx 1$ corresponding
to the $\alpha$-helix (\ref{bc1}).  
\begin{figure}[h]
        \centering
                \includegraphics[width=0.45\textwidth]{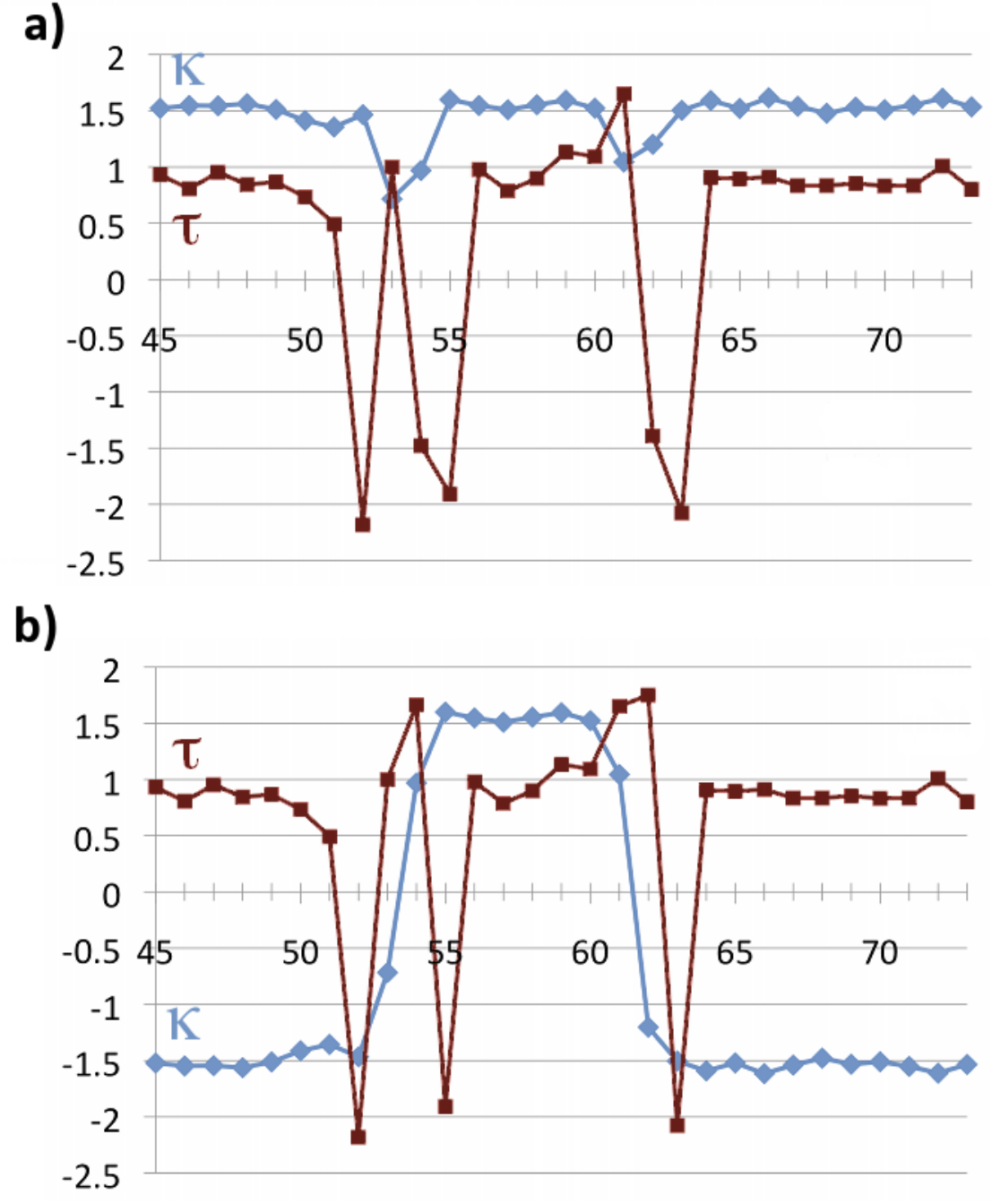}
        \caption{{ 
     (Color online) The profile ofÊ$\kappa_i$Ê(light blue) andÊ$\tau_i$ (dark red)Êalong the 1YRF background. 
     We use PDB indexing of the sites. In a) the $\kappa_i$ are restricted to $[0,\pi]$ 
     and in b) this region is extended to $[-\pi, \pi]$ using (\ref{Z2}). }}
       \label{Figure 4}
\end{figure}
Notice the irregular structure of the torsion angle $\tau_i$ in the loop (soliton) regions. {\it A priori} we expect
from (\ref{tauk}) that the torsion angle should have a regular profile. However, the numerical values that
we compute from (\ref{tauk}) are not restricted to the fundamental range  $\tau_i \in [-\pi, \pi]$, they can
take values beyond this range. The irregular structure of $\tau_i$ follows when we convert  the 
values to the fundamental range, using $2\pi$ periodicity of $\tau_i$ in the discrete Frenet equation (\ref{DFE2}).
Similarly we observe slight irregularity in the $\kappa_i$ profile. This can also be removed
if we allow $\kappa_i$ to take values beyond $[-\pi, \pi]$ and use the $2\pi$ periodicity.  
But  in the case of 1YRF the 
improvement in the precision turns out to be very small, and consequently we 
search for a solution of (\ref{nlse}) by assuming that $\kappa_i \in [-\pi, \pi]$.

In Figure 5 we show the distribution of the side-chain angles ($\theta_i, \varphi_i$) in YRF, by plotting the 
tips of the unit vector (\ref{uvec}) on the two-sphere of Figure 2.  As expected, they are located
in the $\alpha$-helix region of  Figure 2a) except along the loops, where they are located 
outside of the regular structure regions.
\begin{figure}[h]
        \centering
                \includegraphics[width=0.45\textwidth]{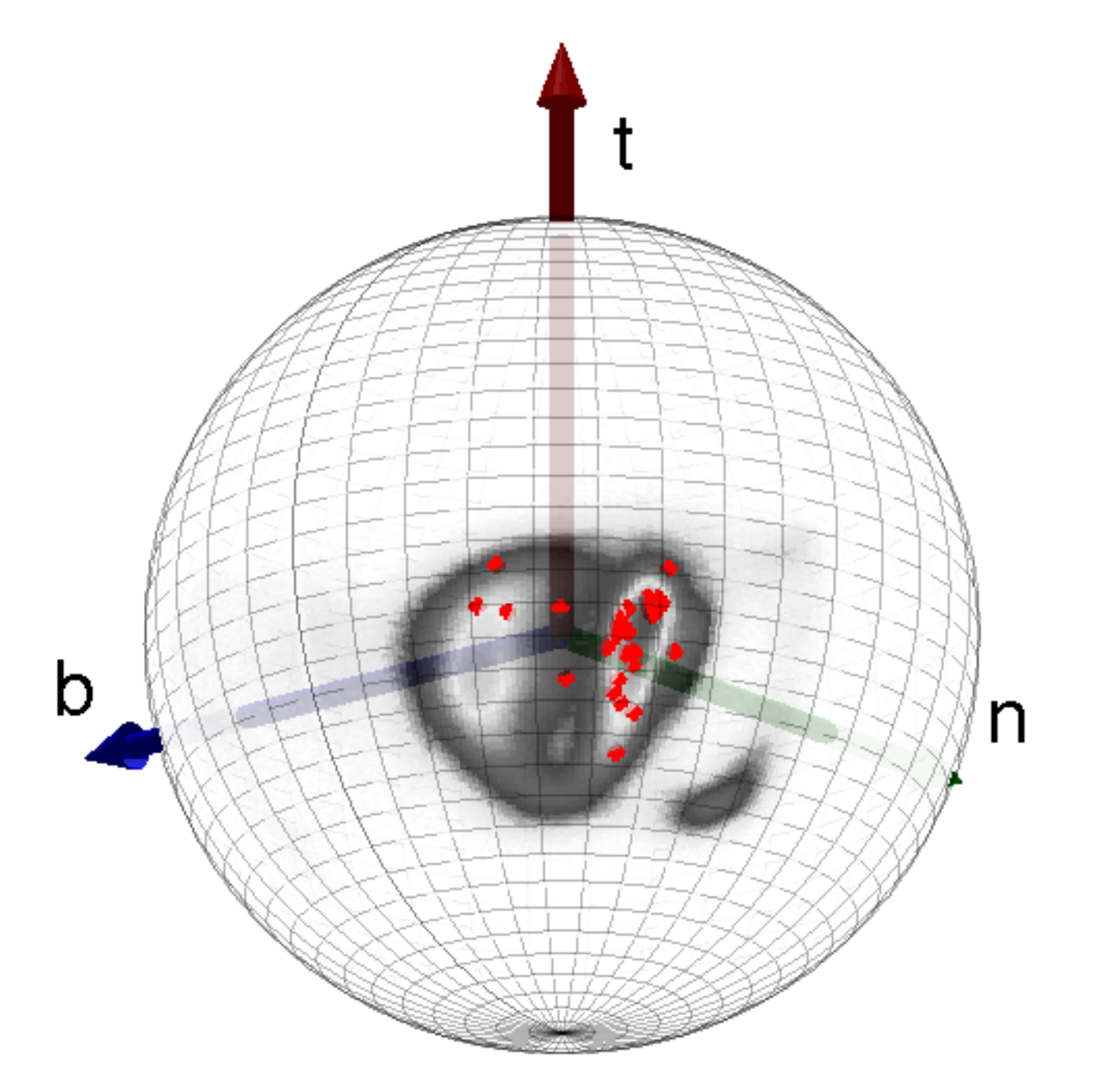}
        \caption{{ 
     (Color online)   The directional distribution of the side-chain angles ($\theta_i, \varphi_i$). The background coincides 
     with the annulus in Figure 2a).  }}
       \label{Figure 5}
\end{figure}

We start by solving the classical equations of motion for $\kappa_i$ from  (\ref{nlse}). We then
construct the remaining variables ($\tau_i, \theta_i, \varphi_i$)  in terms of $\kappa_i$ using 
(\ref{tauk}), (\ref{thei}) and (\ref{phii}); Since the ($\theta_i, \varphi_i$) contributions to the
$\kappa_i$ potential (\ref{Vkappa}) are minuscule, we ignore the corresponding parameters
in constructing the solution to the DNLS equation for $\kappa_i$. 
We use the iterative algorithm and procedure described in 
\cite{her}, \cite{nora}, and our results are summarized in Figure 6 and Table 1.  We have been able to substantially improve the accuracy reported 
in \cite{nora}, in particular for the first soliton. We now reach a RMSD accuracy less than 0.4 \.A even
when we include the
side-chain C$_\beta$ atoms. The result is  clearly within the Debye-Waller fluctuation distance regime that we compute
from the experimental B-factors in the PDB data.
\begin{figure}[h]
        \centering
                \includegraphics[width=0.45\textwidth]{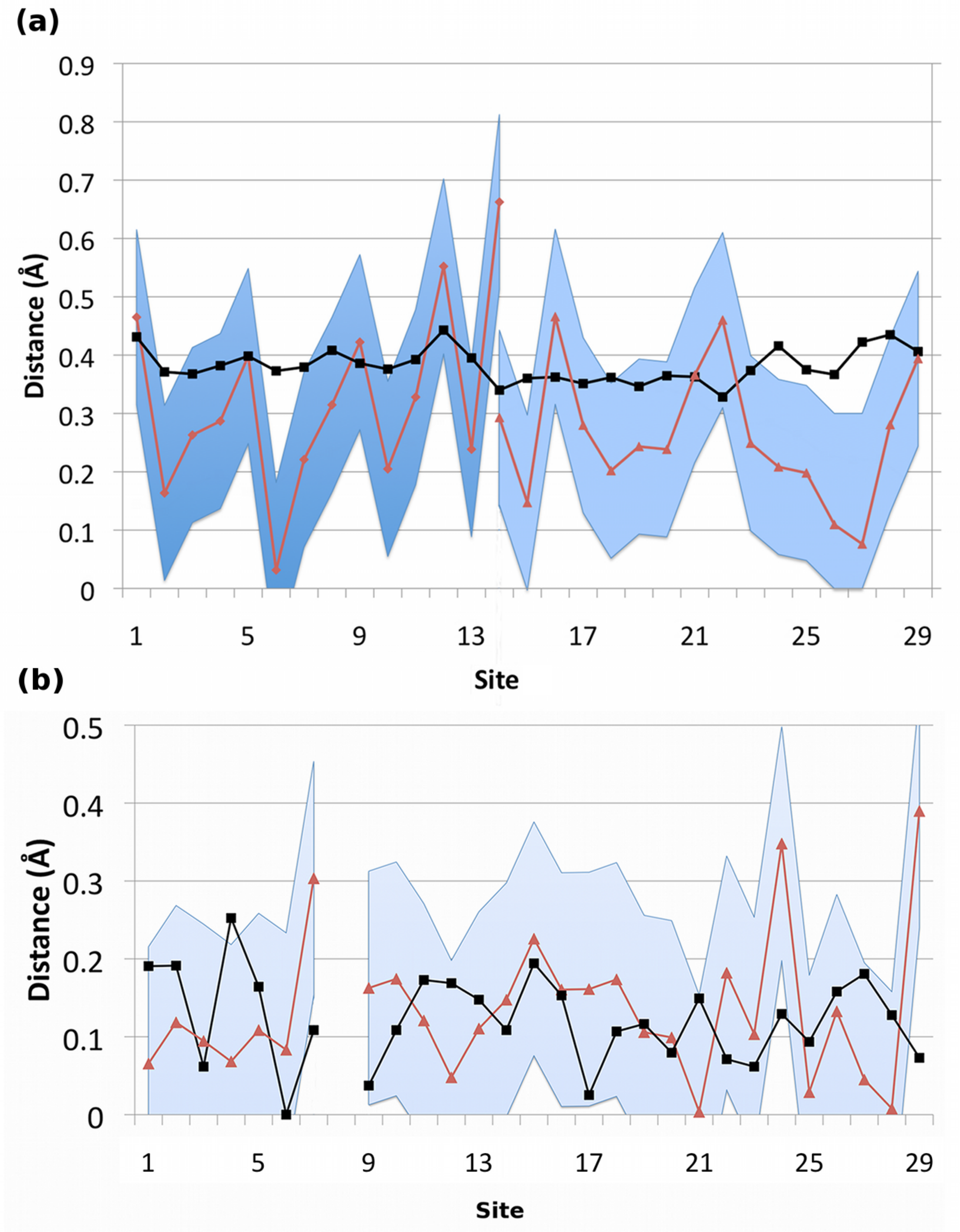}
        \caption{{ 
    (Color online)  Comparison between our soliton solutions in red (gray) and the experimental 
     B-factor fluctuation distance of  PDB data for 1YRF (black) along the backbone.
     In Figure a) for the C$_\alpha$, and in Figure b) for the C$_\beta$ where the experimental accuracy is estimated from (\ref{B}). 
     The shaded region describes the 0.15 \.A  zero point fluctuations around solitons. The cut in Figure a) 
     at sites 13-14 is where
     the two solitons overlap (Phe-58 in PDB), and the empty space in Figure b) is due to glycine that has no C$_\beta$.   }}
       \label{Figure 6}
\end{figure}
{\small
\begin{table}[htb]
       \centering
\begin{tabular}{|c|c|c|}
\hline  parameter & soliton-1 & soliton-2    \tabularnewline
\hline \hline
$q_1$ & 0.459712  &   0.995867 \tabularnewline
\hline
$q_2$   & 4.5533320 &  9.408796  \tabularnewline
\hline
$m_1$ & 1.504535 & 1.550322  \tabularnewline
\hline
$m_2$  & 1.512836  &  1.535081   \tabularnewline
\hline
$a_\tau$ & 9.5752137e-9 & 7.840467e-6   \tabularnewline
\hline
$b_\tau$ & -676965e-11 &  -4.973244e-9  \tabularnewline
\hline
$c_\tau$ & 4.875744e-9 &  4.2733696e-6  \tabularnewline
\hline
$d_\tau$ & -2.917129e-9 &  -2.431388e-6   \tabularnewline
\hline
 $a_\theta$ &1.514770  &  1.322495  \tabularnewline
\hline
$b_\theta$ & -0.0017952 & -0.018619  \tabularnewline
\hline
$d_\theta$ &  0.0420877 &  6.930946e-8  \tabularnewline
\hline
$a_\varphi$ & 0.544859 &   0.3594184 \tabularnewline
\hline
$b_\varphi$ & 5.66111e-5  & 3.83253e-4   \tabularnewline
\hline
$d_\varphi$ & -0.1845828  &  -0.226012  \tabularnewline
\hline  \hline
RMSD (\.A) & 0.38 & 0.32 \tabularnewline
\hline
\end{tabular}
\caption{
Parameter values for the two-soliton solution that describes the two
loops of  1YRF with a combined 0.39\.A accuracy for both
C$_\alpha$ and C$_\beta$ atoms.  The displayed  RMSD values are for the
individual solitons. 
The soliton-1 is located at  Glu-45 - Phe-58 and the soliton-2 is
located at Phe-58 - Lys-73.
We utilize scale invariance to set all $c_\theta = c_\varphi = 1$. The result has
sensitivity to the accuracy of parameters, because a folded protein is a piecewise linear
polygonal chain with a positive Liapunov exponent.
 }
       \label{tab:para}
\end{table}
}


In Figure 6a) we display the distance between the computed and the experimentally measured C$_\alpha$ atoms
(excluding the N and C terminals).  The shaded region in Figure 6a) describes the 0.15\.A zero point fluctuations \cite{peng} around
our solitons. For comparison, we also display the experimental  Debye-Waller B-factor 
fluctuation distances, obtained from the PDB data. Except for the end
point of soliton 1 (residue 58), our soliton solutions describe the backbone well within the limits of experimental accuracy.

In Figure 6b) we present our results for the C$_\beta$ nutation, in comparison with the experimental data. 
We also present an estimate for the experimental
uncertainties that we estimate as follows:  The experimental B-factors 
give an estimate for the {\it absolute} fluctuation distance
around the measured position. 
But now we are interested in estimating the (much smaller) {\it relative} error in the 
position of C$_\beta$ with respect to the position of the ensuing C$_\alpha$. For this 
we introduce the {\it relative}  B-factor
\begin{equation}
B_{rel} \ = \  \left|B_{\alpha} - B_{\beta}\right|
\label{B}
\end{equation}
In Figure 6b) we display the ensuing  fluctuation distances  that we have computed from the Debye-Waller relation using
(\ref{B}) in lieu of 
the B-factor. The precision of our computed results compare well with these  experimental relative B-factor  errors:
For most of the sites the difference is no more than the 0.15\.A estimate for zero point fluctuations.

Finally, Figure 7 shows our soliton solution together with the 1YRF configuration in PDB. 
%
\begin{figure}[h]
        \centering
                \includegraphics[width=0.5\textwidth]{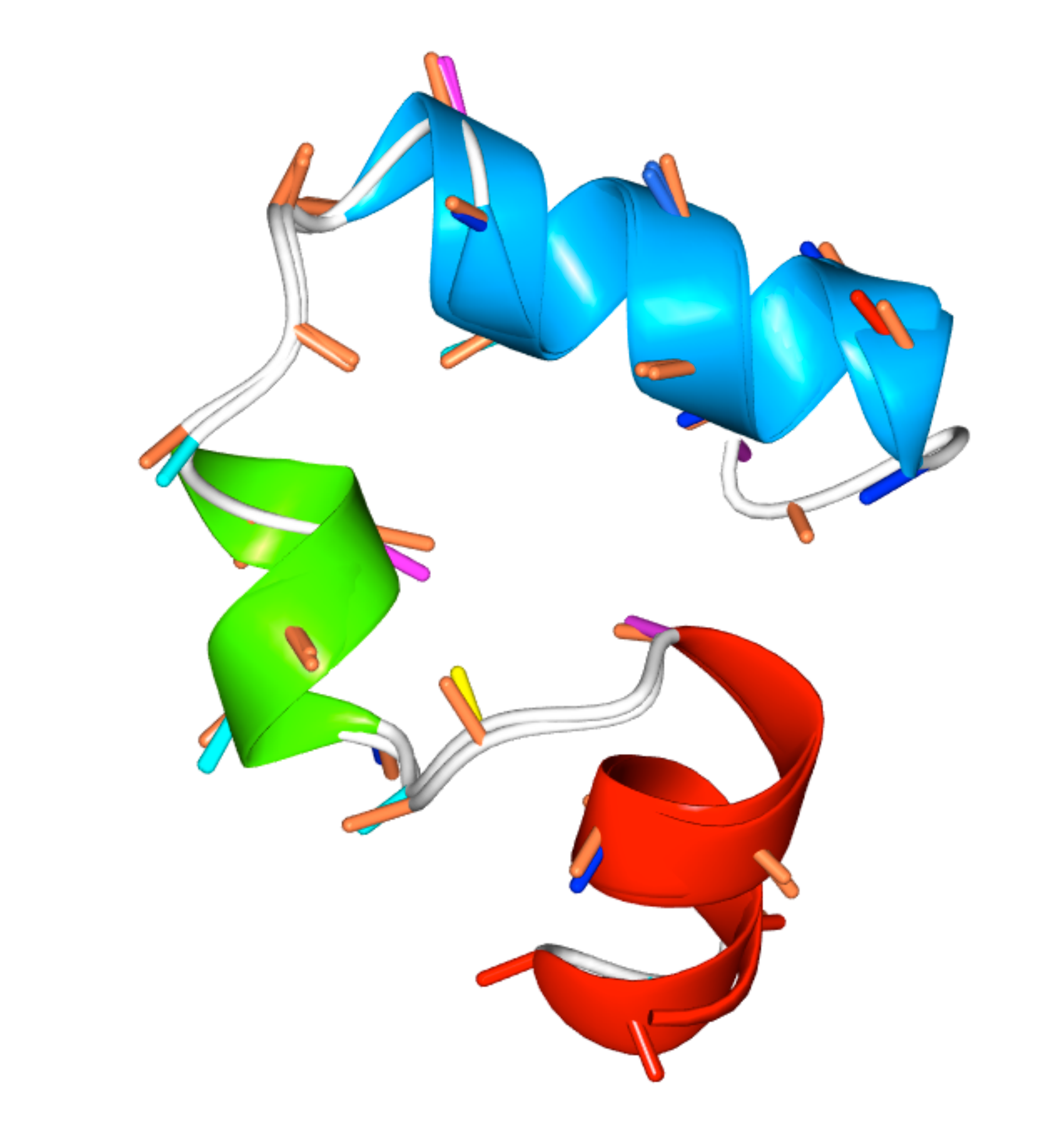}
        \caption{{(Color online) 
    A cartoon comparison of HP35 with our soliton solution summarized in Table 1. The combined C$_\alpha$ and C$_\beta$
    root-mean-square distance is 0.39 \.A which equals  the experimental
     Debye-Waller B-factor fluctuation distance for the central carbons.  }}
       \label{Figure 7 }
\end{figure}

\section{Summary}

In conclusion, the paradigm assumption that the tetrahedral covalent symmetry around the backbone C$_\alpha$ carbons
is transferable, is correct to a good precision. However, with the advent of third and fourth generation
X-ray sources there is now a rapid growth in the number of 
protein structures with sub-{\AA}ngstr\"om resolution. This makes it possible
to scrutinize small corrections to this paradigm. We have found, that the backbone $N-C_\alpha-C$ bond angle shows 
systematic deviations from the ideal value, in a manner that is 
in direct correspondence with the corresponding secondary  structure environment. 
We have investigated how this effect propagates to the orientation of the C$_\beta$ carbon. 
We  have found that the angular orientations of the C$_\beta$ carbon similarly deviate from their ideal
values, in a manner which is in a one-to-one correspondence with the underlying secondary 
structure environment. 

We have presented  a simple energy function that is based on the concept of universality, 
to model the secondary structure dependence in the C$_\beta$ orientations. As an example,  we have 
constructed the C$_\alpha$-C$_\beta$ backbone of HP35 villin, where we reach an accuracy that matches the 
experimental B-factor fluctuation distances. We propose that our observations  and theoretical
proposals could form a basis for  the development of both more
accurate refinement tools for experimental data analysis, and of more precise  
theoretical and computational MD force fields,  to model the atomic level structure and dynamics of folded proteins.

\end{document}